\documentstyle[aps]{revtex}
\begin{document}
\narrowtext

\title{Charge Symmetry Breaking in 500 MeV Nucleon-Trinucleon
        Scattering}
  \author{Tim Mefford and Rubin H. Landau}
\address{Department of Physics, Oregon State University, Corvallis, OR
97331}

\date{\today}
\maketitle

\begin{abstract}
Elastic nucleon scattering from the $\mbox{}^3$He and $\mbox{}^3$H
mirror nuclei is examined as a test of charge symmetry violation. The
differential cross sections at $500$ MeV are calculated using a
microscopic, momentum-space optical potential including the full
coupling of two spin $1/2$ particles and an exact treatment of the
Coulomb force.  The charge-symmetry-breaking effects investigated
arise from a violation within the nuclear structure, from the
p-nucleus Coulomb force, and from the mass differences of the charge
symmetric states. Measurements likely to reveal reliable information
are noted.
\end{abstract}

%\narrowtext

\section{Introduction}

Charge symmetry (CS) is an approximate symmetry of the strong force
arising from the nearly equal masses of the up and down quarks.  At
the nuclear level this means that, apart from electromagnetic and weak
effects, the interactions of nuclear systems must be the same as the
interactions of their charge-symmetric counterparts as long as the
spin-space states involved remain identical.  Yet even if the
nucleon-nucleon strong forces respected this symmetry, in an
experimental measurement we would expect some CS violation to arise
from the Coulomb force, from the different masses for the neutron and
proton, and from the different masses of charge-symmetric (mirror)
nuclei.  If the nucleon-nucleon strong force does not obey
CS\cite{cs.thomas}, we would expect further violation within the
nuclear structure and within the nucleon--nucleus interaction.

A useful approach to testing CS in scattering from the trinucleon
system is to remove some experimental uncertainties by forming ratios
of cross sections which should be identically equal if CS were obeyed.
For the pion-trinucleon system, charge symmetry predicts\cite{landau}
\begin{eqnarray}
	r1 & = &
\frac{d\sigma/d\Omega(\pi^+-\mbox{}^{3}\mbox{H})}{d\sigma/d
\Omega(\pi^--\mbox{}^{3}\mbox{He})} \equiv 1,  \\
	r2 & = &
\frac{d\sigma/d\Omega(\pi^--\mbox{}^{3}\mbox{H})}
	{d\sigma/d\Omega(\pi^+-\mbox{}^{3}\mbox{He})}   \equiv 1.
\end{eqnarray}
When Pillai {\em et al.}\cite{Pillai,nefkens} measured these ratios
they found large deviations from $1$ and concluded that substantial CS
violation must be occurring.  The analysis by Kim {\em et
al.}\cite{Kim} showed that the $\pi$-trinucleon Coulomb force is
largely responsible for these ratios differing from $1$, specifically,
the ratios get relatively large when
$d\sigma/d\Omega(\pi^{\mp}-\mbox{}^{3}\mbox{He})$ have minima, and
these minima are sensitive to Coulomb-nuclear interference. In
addition, Kim {\em et al.} concluded that meaningful predictions of
these ratios required a theory which could reproduce accurately the
individual cross sections, and not just the ratios.

In a further study, Gibbs and Gibson\cite{Gibbs} concluded that the
magnitude of the ratios $r1$ and $r2$ after the $\pi$-trinucleon
Coulomb force is included arises from CS violation within the nuclear
structure. They fit the pion scattering data by introducing
differences in the root-mean-square nuclear radii,
\begin{eqnarray}
	 R_{n}(\mbox{}^{3}\mbox{H}) - R_{p}(\mbox{}^{3}\mbox{He})
	= -0.30 \pm 0.008 \mbox{ fm},  \label{gibbs1}\\
	  R_{n}(\mbox{}^{3}\mbox{He}) - R_{p}(\mbox{}^{3}\mbox{H})
	= 0.035 \pm 0.007 \mbox{ fm},\label{gibbs2}
\end{eqnarray}
differences which would vanish if CS were good within the nuclear
structure\cite{schiff}.  While these differences are relatively small
compared to the nuclear radius of $\sim 2$ fm, and probably close to
the level of uncertainty in strong-interaction calculations, they are
approximately the same size as the CS violation found by including the
Coulomb force in Fadeev calculations of nuclear structure. It is,
accordingly, interesting to see if other hadron probes can confirm
this degree of CS violation.

For the nucleon-trinucleon system, charge symmetry demands equal cross
sections for $p-\mbox{}^{3}$H and $n-\mbox{}^{3}$He reactions, and,
independently, for the $p-\mbox{}^{3}$He and $n-\mbox{}^{3}$H
reactions:
\begin{eqnarray}
	r1 & \stackrel{def}{=} &
	\frac{d\sigma/d\Omega(p-\mbox{}^{3}\mbox{H})}{d\sigma/d
	\Omega(n-\mbox{}^{3}\mbox{He})} \equiv 1, \label{r1}\\
	r2 &
	\stackrel{def}{=} &
	\frac{d\sigma/d\Omega(n-\mbox{}^{3}\mbox{H})}{d\sigma/d
	\Omega(p-\mbox{}^{3}\mbox{He})} \equiv 1, \label{r2}\\
\cal{R} &
	\stackrel{def}{=} & r1 \times r2 \equiv 1,\label{R}
\end{eqnarray}
where $\cal{R}$ is called a ``superratio''.  It is the calculation of
these ratios and the sensitivities of these ratios to several aspects
of CS violation with which we are concerned in this paper.  A direct CS
violation within the nucleon-nucleon interaction\cite{cs.thomas} is
not considered here.

\section{The Calculation}

Our calculation is based on a solution of the Lippmann-Schwinger equation
with a microscopic, nonlocal, momentum-space optical potential
including all spin $\frac{1}{2} \times \frac{1}{2}$ couplings and an
exact treatment of the Coulomb potential\cite{paez,me,lu}. The
potential is the sum of nuclear ($nuc$) and Coulomb ($coul$) parts:
\begin{eqnarray}
	V({\bf k}', {\bf k}, E) &=& V^{nuc}({\bf k}', {\bf k},E) +
        V^{coul} ({\bf k}', {\bf k}) , \\
		V^{nuc}({\bf k',k}, E)
        &\simeq& N
        \left\{(t^{Nn}_{a+b}+t^{Nn}_{e}\vec{\sigma}^{N}_{n})
        \rho^{n}_{mt}(q=|{\bf k'- k}|) \right.  \nonumber \\
	&&+ \left.
	\left[
        t^{Nn}_{a-b} \vec{\sigma}^{N}_{n} \vec{\sigma}^{3}_{n} +
        t^{Nn}_{e} \vec{\sigma}^{3}_{n} + t^{Nn}_{c+d}
        \vec{\sigma}^{N}_{m} \vec{\sigma}^{3}_{m} + t^{Nn}_{c-d}
        \vec{\sigma}^{N}_{l} \vec{\sigma}^{3}_{l} + t^{Nn}_{c+d}
        (\vec{\sigma}^{N}_{m} \vec{\sigma}^{3}_{l} +
        \vec{\sigma}^{N}_{l} \vec{\sigma}^{3}_{m})\right]
        \rho^{n}_{sp}(q) \right\} \nonumber \\ &&+ Z \left\{
        (t^{Np}_{a+b}+t^{Np}_{e}\vec{\sigma}^{N}_{n}) \rho^{n}_{mt}(q)
        \right. \label{U} \\ &&+ \left. \left[t^{Np}_{a-b}
        \vec{\sigma}^{N}_{n} \vec{\sigma}^{3}_{n} + t^{Np}_{e}
        \vec{\sigma}^{3}_{n} + t^{Np}_{c+d} \vec{\sigma}^{N}_{m}
        \vec{\sigma}^{3}_{m} + t^{Np}_{c-d} \vec{\sigma}^{N}_{l}
        \vec{\sigma}^{3}_{l} + t^{Np}_{c+d} ( \vec{\sigma}^{N}_{m}
        \vec{\sigma}^{3}_{l} + \vec{\sigma}^{N}_{l}
        \vec{\sigma}^{3}_{m})\right ] \rho^{n}_{sp}(q) \right\}.
        \nonumber
\end{eqnarray}
Here the $t$'s are elementary, two-nucleon T matrices with their
superscripts indicating the nucleons involved, and with their
subscripts indicating the spin dependences\cite{isospin}. The
$\sigma$'s are Pauli spinors with their superscripts indicating the
beam nucleon and trinucleon target involved, and with their subscripts
indicating $\sigma$'s projections onto the three independent scattering
vectors, $\hat{\bf n}\propto {\bf k \times k'}$, $\hat{\bf
m}\propto{\bf k-k'}$, and $\hat{\bf l} \propto{\bf k+k'}$.  The
$\rho$'s are four independent form factors describing the distribution
of matter ($mt$) and spin ($sp$) within the nucleus.

The nuclear form factors are a key ingredient of the optical potential
and possibly the most interesting path upon which CS violation enters
our calculation.  If CS were good for the trinucleon structure, the form
factors would obey the relation
\begin{equation}
    \rho^{p}_{\alpha}(^{3}\mbox{He}) =
	\rho^{n}_{\alpha}(^{3}\mbox{H}), \ \ \
	\rho^{p}_{\alpha}(^{3}\mbox{H}) =
	\rho^{n}_{\alpha}(^{3}\mbox{He}), \ \ \ (\alpha=mt, \,sp).\label{assume}
\end{equation}
The relations (\ref{assume}) reflect the CS of mirror nuclei: the
distribution of the two ``like'' nucleons is the same in both nuclei,
as is the separate distributions of the ``unlike'' nucleons, and this
is independent of whether the nucleons are neutrons or protons.

If we ignore meson-exchange currents, the matter and spin form factors
are related to the charge ($ch$) and magnetic ($mg$) form factors of the
trinucleon system with the finite proton size removed\cite{paez},
\begin{eqnarray}
   \rho_{mt}^{p}(\mbox{He}) &=& F_{ch}(\mbox{He})/f,\\
	\rho_{mt}^{n}(\mbox{He}) &=& F_{ch}(\mbox{H})/f , \label{13}
	\\ \rho_{sp}^{n}(\mbox{He}) &=& [\mu_{p}^{2}F_{mg}(\mbox{H}) -
	\mu_{n}^{2}F_{mg}(\mbox{He})]/[f(\mu_{p}^{2}-\mu_{n}^{2})],
	\label{14}\\ \rho_{sp}^{p}(\mbox{He}) &=&
	\mu_{p}\mu_{n}[F_{mg}(\mbox{H}) -
	F_{mg}(\mbox{He})]/[2f(\mu_{p}^{2}-\mu_{n}^{2})]. \label{15}
\end{eqnarray}
Here $f$ is the charge form factor of an elementary proton and
$\mu_{p,n}$ the nucleons' magnetic moments. In previous work we have
applied (\ref{13})-({\ref{15}) with realistic charge and magnetic form
factors. For the present calculation, however, we make some
simplifying assumptions which permit a more convenient variation of
nuclear radii and which help eliminate noise from what is already a
numerically challenging calculation. Specifically, we assume that the
distribution of spin  for the unpaired nucleon is the same as its
distribution of matter,
\begin{equation}
	\rho_{sp}^{n}(\mbox{He})
	= \rho_{mt}^{n}(\mbox{He}) ,
\end{equation}
while the spin distribution for the paired nucleons vanish,
\begin{equation}
	\rho_{sp}^{p}(\mbox{He})  = 0.
\end{equation}
To enable convenient variations of the nuclear radii, we use analytic
expressions\cite{Mac} for the form factors.  (Performing the
calculations with numerical, Fadeev form factors\cite{Hadj}, changes
the predictions somewhat, but not the conclusions.)

The nuclear RMS radii used in our calculation are given in
Table~\ref{table1}.  Row one gives the values used assuming charge
symmetry, while rows two and three break CS.  The $1.88$ fm value
derives from the $\mbox{}^{3}$He charge form factor\cite{Mac} and the
$1.70$ fm value from the $\mbox{}^{3}$H charge form
factor\cite{Col,Beck}.  The value of $1.76$ fm for $R_p(^3\mbox{H})$
in row three arises from a recent measurement\cite{Just}, and is
significantly larger than the values of previous measurements. It
would be valuable to have it confirmed with a nucleon probe.

To obtain sufficient numerical accuracy for the large momentum
transfers which occur with protons, we discretized the
coupled-channels Lippmann-Schwinger equation over as many as $64$ grid
points, decomposed the potentials and T matrices into $64$ partial
waves, and included $NN$ partial waves up to $l=4$ (higher NN partial
waves tend to introduce numerical noise). To include the singular,
momentum-space Coulomb potential, we used a cutoff radius of $7$ fm,
and verified that our results are stable for small variations about
this radius\cite{lu}. With all these effects included at the requisite
high precision, the calculation is numerically intensive, and so we
modified the computer code LPOTp to run on a parallel
computer\cite{parallel}.

\section{Results}

The calculations we report here are for $500$ MeV nucleon
scattering. Nearby energies yield similar predictions.  In the top
part of Fig.~\ref{fig1} we show the predicted differential cross
sections for proton and neutron elastic scattering from $\mbox{}^3$He
and $\mbox{}^3$H with all CS-breaking effects included. We see that
the neutron scattering cross sections (dotted and long-dashed curves)
do not exhibit a forward Coulomb peak. But aside from that, the
$p\,\mbox{}^3$He and $n\,\mbox{}^3$H cross sections, and the
$p\,\mbox{}^3$H and $n\,\mbox{}^3$He cross sections, respectively, are
nearly equal (as expected from approximate CS).  We also note that the
proton scattering cross sections (solid and short-dashed curves)
develop distinct oscillations for angles larger than $\sim
100^o$. These oscillations are a consequence of Coulomb-nuclear
interference, and in contrast to pion scattering where the large-angle
ratios all equal one, nucleon oscillations occur because the nuclear
cross section have fallen off to a level comparable to the background
Coulomb cross section.

An important consequence of the backward-hemisphere oscillations in
nucleon-trinucleon scattering is that the ratios (\ref{r1})-(\ref{r2})
deviate considerably from $1$, as we show in the bottom of
Fig.~\ref{fig1}. Unfortunately, while we really do expect this large
an experimental CS-violation signal to occur, we suspect that the
back-angle part is too sensitive to details of Coulomb-nuclear
interference and to uncertainties in the numerical Coulomb procedure\cite{lu}
to
produce reliable information.  Consequently, we consider only
forward-hemisphere measurement of the ratios as reliable, and
henceforth show only them. For example, we notice in the bottom of
Fig.~\ref{fig1} that the forward-hemisphere ratios $r1$ and $r2$
differ smoothly and significantly from $1$. Furthermore, as shown in
Figs.~\ref{fig2}-\ref{fig3}, when the superratio $\cal{R}$ (\ref{R})
is formed, we do obtain a large signal in the forward hemisphere, quite
a bit larger in fact than the one for pion scattering.

In the the top of Fig.~\ref{fig2} we show the predicted superratio
when all CS violation effects are included.  The solid and dashed
curves correspond to the smaller and larger nuclear radii given in
rows 2 and 3 of Table~\ref{table1}. We see that the superratio is
sensitive, but not overly sensitive, to the uncertainty in nuclear
radii. If an experiment could measure the ratio to this level of
precision, an independent measurement of nuclear size should be
possible.

In the the bottom of Fig.~\ref{fig2} we show the predicted superratio
including the CS violation arising from only the (n,p) and
($\mbox{}^3$He,$\mbox{}^3$H) mass differences (solid curve), and from only
the nuclear structure (dashed
curve). The mass difference effect
essentially disappears in comparison to the nuclear structure one. Since a
violation in the nuclear structure  is of more interest than the
mass difference one, this is an encouraging finding.

In the top of Fig.~\ref{fig3} we see the sensitivity of the superratio
$\cal{R}$ to the use of nuclear form factors given by analytic fits to
electron scattering data (solid curve) and given by numerical
solutions to Fadeev equations (dashed curve). In the top part of this
figure, CS violation from only the mass differences is included, and
in the bottom violation from only the p-nucleus Coulomb force is
included.  This figure shows that the superratio in this angular
region is not sensitive to details of the nuclear form factors, but is
sensitive to the nuclear size.  In addition, by comparing
Figs.~\ref{fig2} and \ref{fig3} we see the main CS violation arises
from the nuclear structure, with a somewhat smaller violation arising
from the p-nucleus Coulomb force.  The Coulomb and structure effects
are seen to combine in the $40^o$ region in Fig.~\ref{fig2} to produce
a large effect.

\section{Conclusion}

We have calculated the ratios of differential cross sections for $500$
MeV proton and neutron elastic scattering from $\mbox{}^3$He and
$\mbox{}^3$H. We used a microscopic, momentum-space optical potential
and included the Coulomb force and all spin couplings exactly.  We found
that at large angles the utility of these ratios as a measure of
charge symmetry breaking is low due to highly sensitive
Coulomb-nuclear interference. However, the forward-hemisphere ratio appears
to be quite reliable yet still sensitive to the important CS violation
mechanisms.

We predict that
most of the CS violation in nucleon-trinucleon scattering should arise
from CS violation at the nuclear structure level, with about $1/3$ of
the effect arising from the proton-nucleus Coulomb interaction.  A
measurement of the ratios of cross sections for nucleon-trinucleon
scattering at the $10\%$ level in the forward hemisphere would be a
valuable adjoint to the analogous pion measurements.

\acknowledgements%------------------------------------

We gratefully acknowledge support from the U.S. Department of Energy
under Grant DE-FG06-86ER40283 at Oregon State University.  Some of
the computations were conducted using resources of the Cornell Theory
Center, which receives major funding from the National Science
Foundation and New York State with additional support from Advanced
Research Projects Agency, the National Center for Research Resources
at the National Institutes of Health, IBM Corporation and members of
the Corporate Research Institute.  We also thank Cherri Pancake for
helpful conversations.

%REFERENCES--------------------------------------------

%Figure captions%=========================================

\begin{figure}
\caption{Top: Differential cross-sections for p and n scattering from
$\mbox{}^3$He and $\mbox{}^3$H at $500$ MeV as a function of CM
scattering angle. Bottom: The ratios $r1$ and $r2$ given by equations
(\protect{\ref{r1}}) and (\protect{\ref{r2}}).
All CS-violating effects are included in this figure.}
\label{fig1}
\end{figure}

\begin{figure}
\caption{Top: The superratio $R$, (\protect{\ref{r2}}), with all CS
violating effects included. The solid and dashed curves correspond to
smaller and large nuclear radii given in lines 2 and 3 respectively of
Table~\protect{\ref{table1}}. Bottom: The superratio $R$ including the
CS violation arising from only the (n,p) and
($\mbox{}^3$He,$\mbox{}^3$H) mass differences (solid curve), and from
CS violation arising from only the nuclear structure (dashed curve).}
\label{fig2}
\end{figure}

\begin{figure}
\caption{Top: The sensitivity of the superratio $R$  to use of nuclear form
factors given by analytic fits to electron scattering data (solid
curve) and given by numerical solutions to Fadeev equations (dashed
curve). In this case, CS violation from only the mass differences are
included. Bottom: Same as on top, only now  CS violation
arises from only the p-nucleus Coulomb force.}
\label{fig3}
\end{figure}

\begin{table}\begin{center}\caption{The rms radii of the matter
distributions for the trinucleon system. The first row are the charge
symmetric values, the others include CS breaking.}
\begin{tabular}{l|cccc}
          &   $R_{p} (\mbox{}^{3}\mbox{H})$  & $R_{n} (\mbox{}^{3}\mbox{H})$  &
          $R_{p} (\mbox{}^{3}\mbox{He})$ & $R_{n} (\mbox{}^{3}\mbox{He})$ \\
\hline \hline
CS   &  1.700  &  1.880  &  1.880  &  1.700  \\
CSB1 &  1.700 &  1.850  &  1.880  &  1.735  \\
CSB2 &  1.760 &  1.850  &  1.880  &  1.795  \\
\end{tabular}
\label{table1}\end{center}
\end{table}

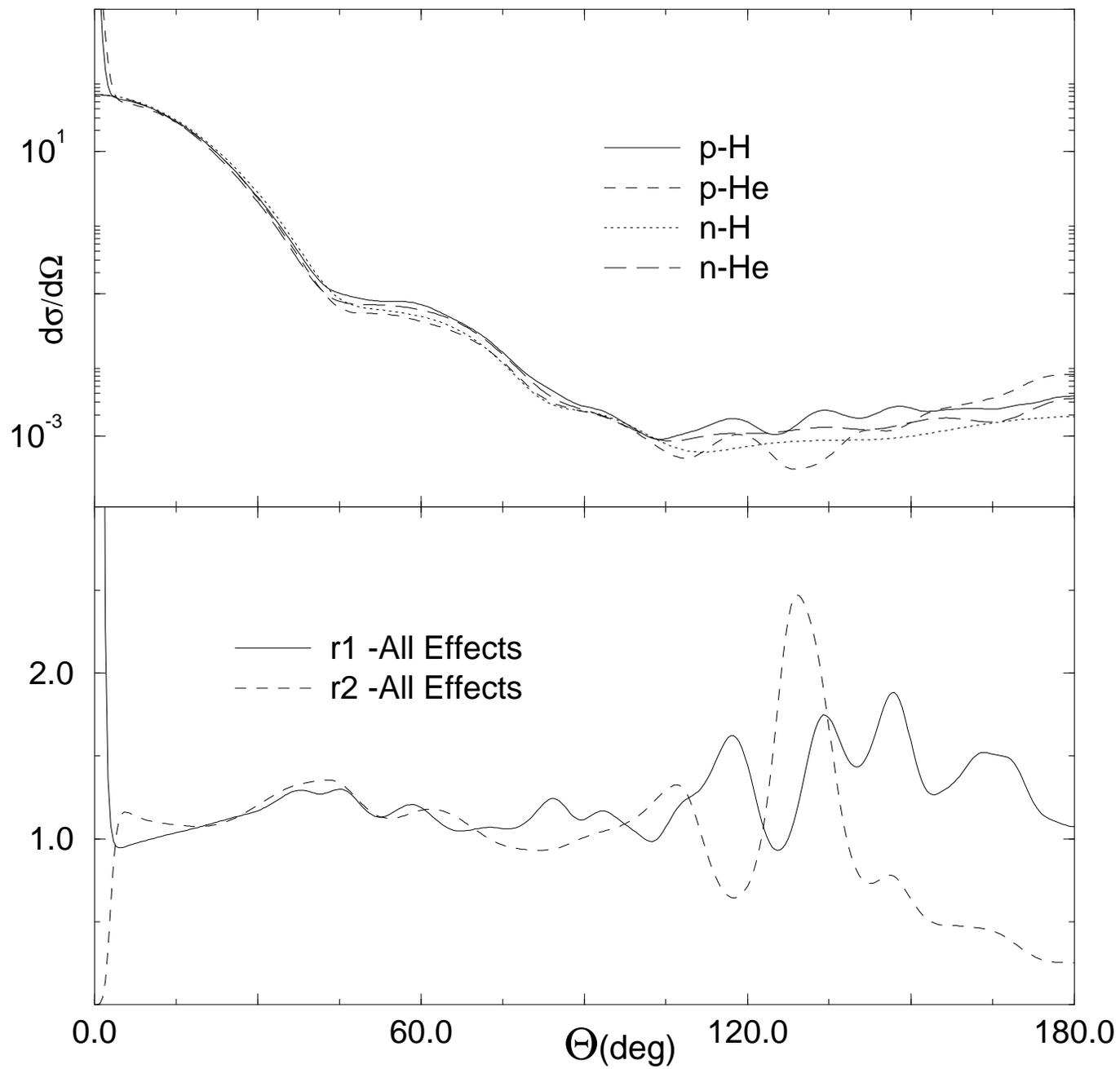
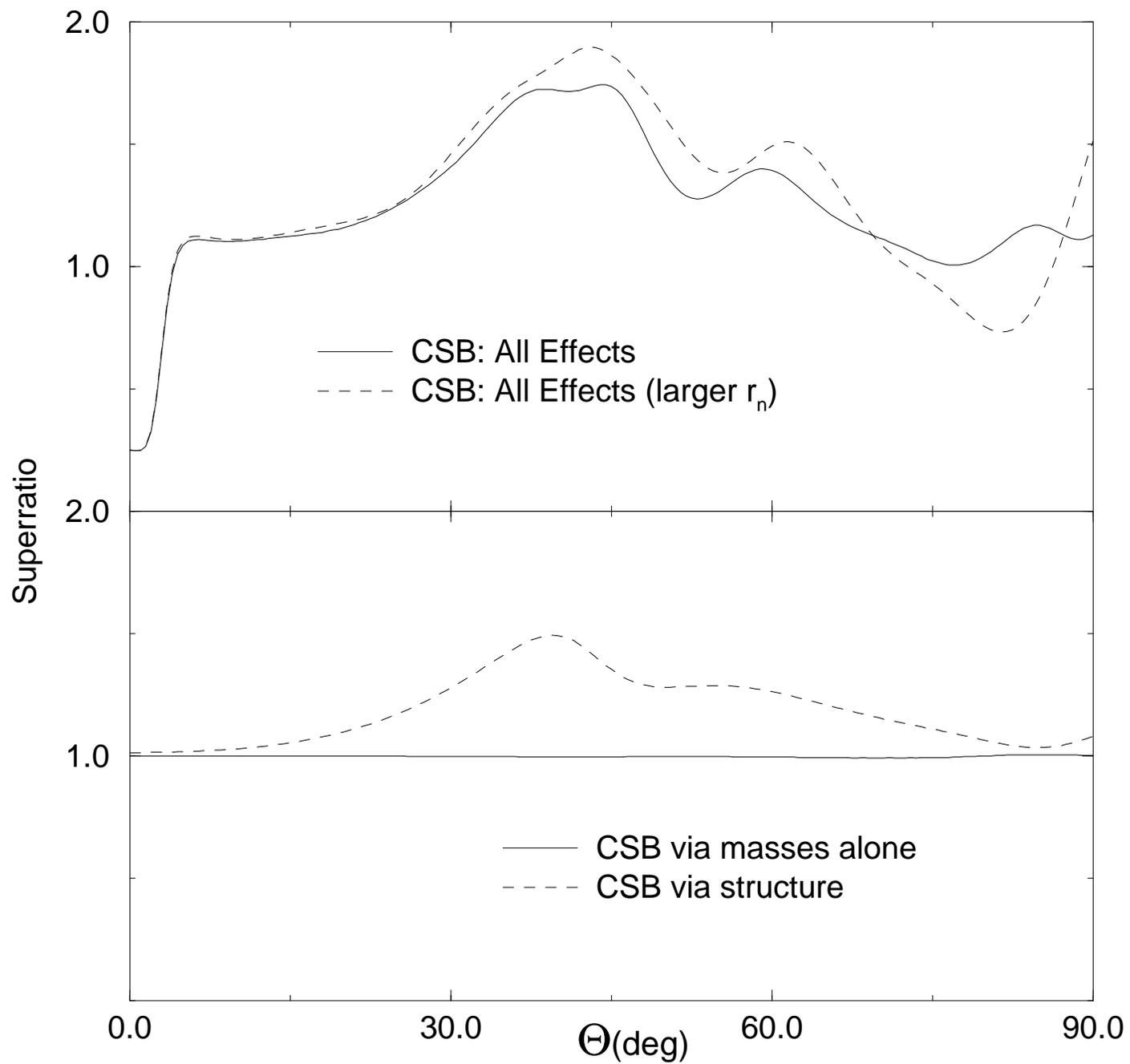
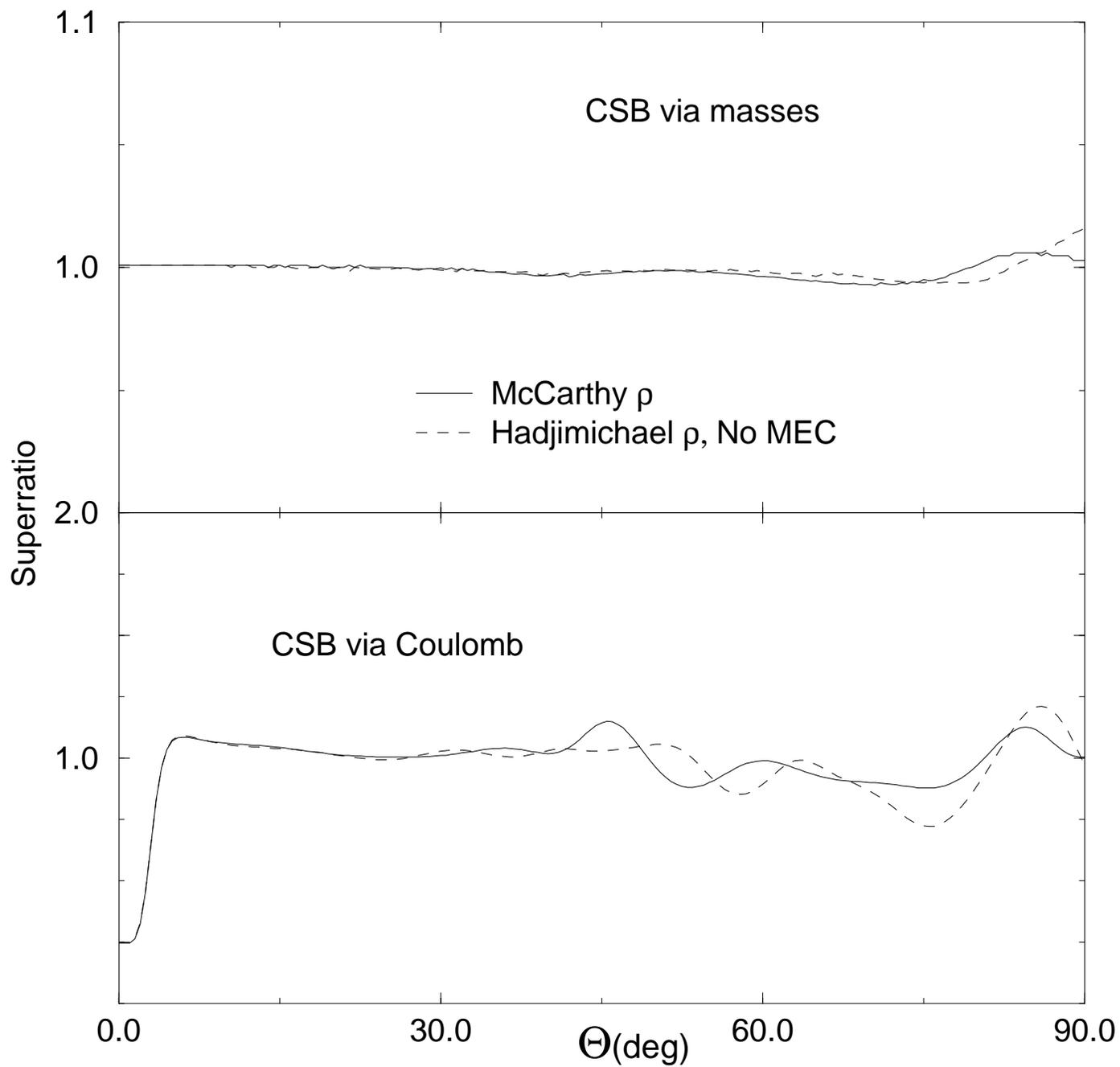
\end{document}